\documentclass{llncs}

\usepackage{times}
\usepackage{algorithm}  
\usepackage[noend]{algpseudocode} 
\usepackage{amsmath}            
\usepackage{amssymb}            
\usepackage{centernot}          
\usepackage{color}              
\usepackage[export]{adjustbox}
\usepackage{graphicx}           
\usepackage{listings}           
\usepackage{makeidx}            
\usepackage{mathtools}          
\usepackage{parcolumns}         
\usepackage{soul}               
\usepackage{verbatim}           

\usepackage{stmaryrd}
\usepackage{paralist}
\usepackage{tabto}

\usepackage{hyperref}
\usepackage{cleveref}

\newcommand{\longversion}[1]{}

\newcommand{\scode}[1]{{\small \texttt{#1}}}

\newcommand{\act}{cmd}

\newcommand{\PC}{\scode{pc}}
\newcommand{\ENVMOVE}{\scode{env\_move}}
\newcommand{\TSTART}{\scode{try\_start}}

\newcommand{\seqcomma}[3]{{#1}_{#2},\dots,{#1}_{#3}}
\newcommand{\seqarrowcheck}[2]{\ifx&#1& \xrightarrow{} \else \xrightarrow{#1^{#2}} \fi }

\newcommand{\reachenv}{Reach_{E}}

\newcommand{\gen}{Gen_E}

\newcommand{\seqarrow}[6]{{#1}_{#3} \seqarrowcheck{#2}{#4} {#1}_{#4} \seqarrowcheck{#2}{#5} \dots \seqarrowcheck{#2}{#6} {#1}_{#6}}
\newcommand{\seqarrowdouble}[4]{{(#1_{#3},#2_{#3})} \xrightarrow{}  \dots \xrightarrow{} (#1_{#4},#2_{#4})}

\newcommand{\val}{\sigma}
\newcommand{\TMAIN}{$t_{M}$}
\newcommand{\TENV}{$t_{E}$}
\newcommand{\PCMAIN}{\PC_M}
\newcommand{\PCENV}{\PC_E}
\newcommand{\VMAIN}{V_M}
\newcommand{\VENV}{V_E}
\newcommand{\VORIG}{V}
\newcommand{\COMP}{\rho}

\newcommand{\PAMAIN}{P_M}
\newcommand{\PAENV}{P_E}
\newcommand{\VAMAIN}{\widehat{\VMAIN}}
\newcommand{\VAENV}{\widehat{\VENV}}

\newcommand{\VORIGANDPC}{V \cup \{\PCENV\}}

\newcommand{\FALSE}{\textit{FALSE}}
\renewcommand{\implies}{\Rightarrow}

\newtheorem{defn}{Definition}
\newtheorem{exmpl}[defn]{Example}

\newif\ifnoappendix
\noappendixtrue

\Crefname{defn}{Def.}{Def.}
\Crefname{exmpl}{Example}{Example}
\Crefname{figure}{Fig.}{Fig.}
\Crefname{section}{Sec.}{Sec.}

\begin{document}

\pagestyle{headings}  

\mainmatter              

\title{Modular Verification of Concurrent Programs via Sequential Model Checking}
\author{Dan Rasin\inst{1} \and Orna Grumberg\inst{1} \and Sharon Shoham\inst{2}}
\institute{Technion -- Israel Institute of Technology
\and
Tel-Aviv University
}

\maketitle              

\begin{abstract}

This work utilizes the plethora of work on verification of sequential programs for the purpose of verifying concurrent programs. We
reduce the verification of a concurrent program to a series of verification tasks of sequential programs.
Our approach is modular in the sense that each sequential verification task roughly corresponds to the verification of a single thread, with some additional information about the environment in which it operates. Information regarding the environment is gathered during the run of the algorithm, by need.

While our approach is general, it specializes on concurrent programs where the threads are structured hierarchically. The idea is to exploit the hierarchy in order to minimize the amount of information that needs to be transferred between threads. To that end, we verify one of the threads, considered ``main'', as a sequential program. Its verification process initiates queries to its ``environment'' (which may contain multiple threads). Those queries are answered by sequential verification, if the environment consists of a single thread, or, otherwise, by applying the same hierarchical algorithm on the environment.

Our technique is fully automatic, and allows us to use any off-the-shelf sequential model checker. We implemented our technique in a tool called CoMuS and evaluated it against established tools for concurrent verification.
Our experiments show that it works particularly well on hierarchically structured programs.
\end{abstract}

\section{Introduction}

Verification of concurrent programs is known to be extremely hard. On top of the challenges inherent in verifying sequential programs, it adds the need to consider a high (typically unbounded) number of thread interleavings.
An appealing direction is to exploit the modular structure of such programs in verification.
Usually, however, a property of the whole system cannot be partitioned into a set of properties that are local to the individual threads.
Thus, some knowledge about the interaction of a thread with its environment is required.

In this work we develop a new approach, which utilizes
the plethora of work on
verification of sequential programs for the purpose of {\em modularly} verifying the safety of concurrent programs. Our technique automatically reduces the verification of a concurrent program to a series of verification tasks of sequential programs.
This allows us to 
benefit from any past, as well as future, progress in
sequential verification.

Our approach is modular in the sense that each sequential verification task roughly corresponds to the verification of a single thread, with some additional information about the environment in which it operates.
This information is \emph{automatically} and \emph{lazily} discovered during the run of the algorithm, when needed.

While our approach is general, it specializes on concurrent programs where the threads are structured hierarchically as it takes a \emph{hierarchical view} of the program.
Namely, for the purpose of verification, one of the threads, \TMAIN{}, is considered ``main'', and all other threads are considered its ``environment''.
The idea is to exploit the hierarchy in order to minimize the amount of information that needs to be transferred between the verification tasks of different threads.

We first analyze \TMAIN{} using sequential verification, where, for soundness,
all interferences from the environment are abstracted (over-approximated) by a function
\ENVMOVE{}, which is called by \TMAIN{} whenever a context switch should be considered. 
Initially, \ENVMOVE{} havocs all shared variables; it is gradually refined during the run of the algorithm. 
When the sequential model checker discovers a violation of safety in \TMAIN{}, it also returns a path leading to the violation.
The path may include calls to \ENVMOVE{}, in which case the violation may be spurious (due to the over-approximation).
Therefore, the algorithm initiates \emph{queries} to the environment of \TMAIN{} whose goal is to check whether \emph{certain} interferences, as observed on the violating path, are feasible\longversion{in the environment}.
Whenever an interference turns out to be infeasible, the \ENVMOVE{} function is refined to exclude it.
Eventually, \ENVMOVE{} becomes precise enough to enable full verification of the desired property on the augmented \TMAIN{}. Alternatively, it can reveal a real \longversion{(non-spurious) }counterexample in \TMAIN{}.

The queries are checked on the environment (that may consist of multiple threads) in the same modular manner. Thus we obtain a {\em hierarchical modular verification}.
Along the algorithm, each thread learns about the next threads in the hierarchy, and is provided with assumptions from former threads in the hierarchy to guide its learning.
When the program has a hierarchical structure that is aligned with the verification process, this makes the assumptions simpler and speeds up verification.

Our technique is fully automatic and performs \emph{unbounded verification}, i.e., it can both find bugs and prove safety in concurrent programs even with unbounded executions (e.g., due to loops), as long as the number of threads is fixed.
It works on the level of program code and generates standard sequential programs in its intermediate checks. This allows us to use any off-the-shelf sequential model-checker.
In particular, we can handle concurrent programs with an infinite state-space, provided that the sequential model checker supports such programs (as is the case in our implementation).

We implemented our technique in a prototype called {\bf Co}ncurrent to {\bf Mu}ltiple {\bf S}equential (CoMuS) and evaluated it against established tools for unbounded verification of concurrent C programs. We use SeaHorn~\cite{DBLP:conf/cav/GurfinkelKKN15} to model check sequential programs.

Our experiments show that the approach works particularly well on programs in which the threads
are arranged as a chain, $t_1, t_2, \ldots, t_k$, where thread $t_i$ depends only on its immediate successor $t_{i+1}$ in the chain. This induces a natural hierarchical structure in which $t_1$ is the main thread with environment $t_2, \ldots, t_k$; thread $t_2$ is the main thread in the environment, and so on.
This structure often occurs in concurrent implementations of dynamic programming algorithms.

To summarize, the main contributions of our work are as follows:
\begin{itemize}
	
\item
We present a new modular verification approach that reduces the verification of a concurrent program to a series of verification tasks of sequential programs. Any off-the-shelf model checker for sequential programs can thus be used.
\item
Our approach takes a hierarchical view of the program,  
where each thread learns about the next threads in the hierarchy, and is provided with assumptions
from former threads to guide its learning.
\item
The needed information on a thread's environment is gathered in the code, automatically and lazily, during the run of the algorithm.
\item
We implemented our approach and showed that as the number of threads grows, it outperforms existing tools on programs that have a hierarchical structure, such as concurrent implementations of dynamic programming algorithms.
\end{itemize}

\subsection{Related Work}

The idea of code transformation to a sequential program appeared in~\cite{DBLP:conf/tacas/TomascoI0TP15a,DBLP:conf/fmcad/TomascoNI0TP16,DBLP:conf/atva/Nguyen0TP16}. However, these works translate the concurrent program to a single nondeterministic sequential program. In contrast, our technique exploits the modular structure of the program.

In the rest of this section, we address unbounded modular techniques for proving safety properties of concurrent programs.
Other techniques use bounded model checking, where the bound can address different parameters, such as the number of context switches~\cite{DBLP:conf/cav/LalR08,DBLP:conf/fmcad/TomascoNI0TP16}, \longversion{the number of }write operations~\cite{DBLP:conf/tacas/TomascoI0TP15a} or \longversion{the number of }loop iterations~\cite{DBLP:conf/cav/RabinovitzG05,akt2013-cav,DBLP:conf/tacas/ZhengELGRDS16}.

The work most closely related to ours is \cite{DBLP:conf/popl/GuptaPR11,DBLP:conf/cav/GuptaPR11}. Their technique uses predicate abstraction of both states and environment transitions (similar to our \ENVMOVE{}), as part of an automatic modular verification framework. The technique also iteratively refines this abstraction by checking possible witnesses of errors. However, they treat all threads symmetrically, whereas our approach exploits a hierarchical view of the program.
\longversion{
In particular, we single out a ``main'' thread from its ``environment'' and initiate queries to learn additional neccessary information. The environment threads are also treated hierarchically, in a similar manner.}%
In addition, \cite{DBLP:conf/popl/GuptaPR11,DBLP:conf/cav/GuptaPR11} explore abstract single threads using reachability trees, which are inherent to their technique. We, on the other hand, represent threads (augmented with some environment information) as stand-alone C programs. Thus, we can use any off-the-shelf model checker to address the ``sequential part'' of the verification problem.

The works in~\cite{DBLP:conf/esop/FlanaganFQ02,DBLP:conf/esop/LeinoM09,DBLP:conf/concur/GavranNKMV15} suggest to apply rely-guarantee reasoning for concurrent (or asynchronous) programs, while the different sections of the program can be verified sequentially. However, their technique requires human effort to specify the rely-guarantee conditions, whereas our approach is completely automatic.

\cite{DBLP:conf/spin/FlanaganQ03} suggests a modular algorithm with rely-guarantee reasoning and automatic condition inference. \cite{DBLP:conf/ictac/MalkisPR06} formalizes the algorithm in the framework of abstract interpretation. However, their algorithm requires finite state systems, and its inferred conditions only refer to changes in global variables. Hence, they fail to prove properties where local variables are necessary for the proof. 
In our approach, reasoning about local variables is allowed, when we learn that they are necessary for verification. Such variables are then turned into global variables, but their behavior is abstracted,
preserving modularity. \cite{DBLP:conf/cav/CohenN07} also tackles the incompetence of modular proofs by exposing local variables as global, according to counterexamples. However, their approach uses BDDs and suits finite state systems.
Similar to \cite{DBLP:conf/popl/GuptaPR11}, they \longversion{also }treat threads symmetrically.
Our approach is applicable to infinite state systems and uses a guided search to derive cross-thread information.

Our queries resemble queries in learning-based compositional  verification~\cite{DBLP:conf/tacas/CobleighGP03,DBLP:journals/fmsd/PasareanuGBCB08}, which are also answered by a model checker. Our hierarchical recursive approach resembles the n-way decomposition handled in~\cite{DBLP:journals/fmsd/PasareanuGBCB08}. However, these works represent programs, assumptions and specification as LTSs,
and although extended to deal with shared memory in \cite{DBLP:conf/cav/SinhaC07} these algorithms are suitable for finite state systems.

Several works such as~\cite{DBLP:journals/entcs/FlanaganQ03,DBLP:conf/spin/GuetaFYS07,DBLP:conf/fmcad/WachterKO13,DBLP:conf/fmcad/PopeeaRW14}, tackle the interleaving explosion problem by performing a thread interleaving reduction. \cite{DBLP:conf/fmcad/WachterKO13} combines partial order reduction \cite{DBLP:books/sp/Godefroid96} with the impact algorithm \cite{DBLP:conf/cav/McMillan06}, whereas \cite{DBLP:conf/fmcad/PopeeaRW14} identifies reducible blocks for compositional verification.  These approaches are complementary to ours, as our first step is performing an interleaving reduction (to identify cut-points for \ENVMOVE{} calls).

\section{Preliminaries} \label{sec:prelim}

\paragraph{Sequential Programs.} A \emph{sequential program} $P$ is
defined by a control flow graph whose nodes are a set of program locations $L$ (also called labels), and whose edges $E$ are a subset of $L \times L$. The program has an initial label, denoted $l^{init} \in L$. Each node $l$ is associated with a command $c \in cmds$, denoted $\act(l)$, which can be an assignment or an if command, as well as \scode{havoc}, \scode{assume} and \scode{assert} (explained below). Intuitively, we think of standard C programs (that may contain loops as well), which can be trivially compiled to such control flow graphs. The program may also include non-recursive functions, which will be handled by inlining.

The program is defined over a set of variables $V$. Conditions in the program are quantifier-free first-order-logic formulas over $V$.
A special variable $\PC\not\in V$, ranging over $L$, indicates the program location.
A \textit{state} $s$ of $P$ is a pair $(l,\val)$ where $l \in L$ is the value of $\PC$ and $\val$ is a valuation of $V$. Variables may have unbounded domains, resulting in a potentially infinite state-space. We also assume the existence of a special error state, denoted $\epsilon=(l_\epsilon,\bot)$.
We denote by $l(s)$ and $\val(s)$ the first and second components (resp.) of a state $s=(l,\val)$.
Given an initialization formula $\phi_{init}$ over $V$, the set of \textit{initial states} consists of all states $(l^{init},\val)$ where $\val \models \phi_{init}$.

For $s=(l,\val)$, let $\act(s)=\act(l)$. We denote $next(s)=\{s' \mid s'$ can be obtained from $s$ using $\act(s)\}$.
This set is defined according to the command.
In particular,
$s' \in next(s)$ implies that $(l(s), l(s')) \in E$.
The definition of $next(s)$ for assignments and if commands is standard. A \scode{$v$=havoc()} command assigns a non-deterministic value to the variable $v$.
An \scode{assume($b$)} command is used to disregard any computation in which the condition $b$ does not hold.
\longversion{It does nothing, otherwise.}%
Formally, if $s=(l,\val)$ and $\act(s)$=\scode{assume($b$)}, 
then $\val \vDash b \implies next(s)=\{ (l',\val) \}$ where $l' \neq l_\epsilon$ is the unique label such that $(l,l') \in E$, and $\val \nvDash b \implies next(s)=\emptyset$.
An \scode{assert($b$)} command is defined similarly, except that it moves to the error state if $b$ is violated. \longversion{That is,  $\val \nvDash b \implies next(s)=\{ e \}$.}

A \textit{computation} $\COMP$ of $P$ is a sequence $\COMP=\seqarrow{s}{}{0}{1}{2}{n}$ for some $n \geq 0$ s.t. for every two adjacent states $s_i,s_{i+1}$: $s_{i+1} \in next(s_i)$. $\COMP$ is an \textit{initial computation} in $P$ if it starts from an initial state. $\COMP$ is a \textit{reachable computation} in $P$ if there exists an initial computation $\COMP^{'}$ for which $\COMP$ is the suffix.
The \textit{path} of a computation $\seqarrowdouble{l}{\val}{0}{n}$ is the sequence of program locations $\seqcomma{l}{0}{n}$.

\paragraph{Preconditions and Postconditions.} Given a condition $q$ over 
$V$ and an edge $e=(l,l')$, a \emph{precondition} of $q$ w.r.t. $e$, denoted $pre(e, q)$, is any condition $p$ such that for every state $s$, if $\val(s) \vDash p$ and $l(s)=l$ then there exists $s'\in next(s)$ s.t. $\val(s') \vDash q$ and $l(s')=l'$\footnote{Note that our definition of a precondition does not require all the successors to satisfy $q$.}. A precondition extends to a path $\pi=\seqcomma{l}{0}{n}$
in the natural way. 
The \emph{weakest precondition} of $q$ w.r.t. $e$ (resp., $\pi$) is implied by any other precondition, and can be computed in the standard way~\cite{Dijkstra76}. We denote it $wp(e, q)$ (resp., $wp(\pi, q)$).

A postcondition of $p$ w.r.t $e=(l,l')$, denoted $post(e, p)$, is any condition $q$ such that if $\val(s) \vDash p$, $l(s) = l$ then for every $s' \in next(s)$, if $l(s') = l'$ then $\val(s') \vDash q$.
Postconditions can also be extended to paths $\pi=\seqcomma{l}{0}{n}$. We use $post(\pi, p)$ to denote a postcondition of condition $p$ w.r.t. path $\pi$.

\paragraph{Concurrent Programs} A \textit{concurrent program} $P$ consists of multiple threads $\seqcomma{t}{1}{m}$, where each \textit{thread} $t_i$ has the same syntax as a sequential program over a set of variables $V_i$ and a program location variable \scode{pc$_i$}.
The threads communicate through shared variables, meaning that generally $V_i,V_j$ are not disjoint for $i \neq j$.
A variable is \textit{written} by $t_i$ if it appears on the left hand side of any assignment in $t_i$. A variable $v$ is \textit{shared} between two threads $t_i$, $t_j$ if $v \in V_i \cap V_j$. A variable $v \in V_i$ is a \textit{local} variable of $t_i$ if $v \not\in V_j$ for every $j \neq i$.
Let $V = \bigcup_{i=1}^{m} V_i$. A \textit{state} of \longversion{the concurrent program }$P$ is a pair $(\overline{l}, \val)$, where $\val$ is a valuation of $V$ and $\overline{l}=(\seqcomma{l}{1}{m})$ where $l_i$ is the value of $\PC_i$. We also assume one common error state $\epsilon$. Given an initialization formula $\phi_{init}$ over $V$, the set of \textit{initial states} consists of all states $(\overline{l}^{init},\val)$ where $\val \vDash \phi_{init}$ and $l^{init}_i$ is the initial label of $t_i$.

The execution of a concurrent program is interleaving, meaning that exactly one thread performs a command at each step, and the next thread to execute is chosen non-deterministically.
We consider a sequentially consistent semantics in which the effect of a single command on the memory is immediate.
For $s = (\overline{l}, \val)$, let $\act(s, t_i)$ denote the command of thread $t_i$ at label $l_i$.
We denote $next(s,t_i)=\{ s' \mid s'$ can be obtained from $s$ after $t_i$ performs $\act(s,t_i)\}$.
A \textit{computation} $\COMP$ of the concurrent program $P$ is a sequence $\seqarrow{s}{t}{0}{1}{2}{n}$ s.t. for every two adjacent states $s_i,s_{i+1}$: $s_{i+1} \in next(s_i, t^{i+1})$.
We say that $\COMP$ is a computation of thread $t$ in $P$ if $t^j=t$ for every $1 \leq j \leq n$. We define \emph{initial} and \emph{reachable} computations as in the sequential case, but w.r.t. computations of the concurrent program.

We 
support 
synchronization operations by modeling them with atomic control commands. For example, \scode{Lock(lock)} is modeled by atomic execution of \scode{assume(lock = false); lock = true}. Since our technique models context switches by explicit calls to \ENVMOVE{},
we are able to prevent context switches between these commands.

\paragraph{Safety.} A computation of a (sequential or concurrent) program is \emph{violating} if it ends in the error state \longversion{(i.e., the last step of the computation executes the command \scode{assert($b$) }at a state $s$ such that $\val(s) \nvDash b$)}. The computation is \emph{safe} otherwise. A (sequential or concurrent) program is \emph{safe} if it has no initial violating computations.
In the case of a sequential program, we refer to the path of a violating computation as a \emph{violating path}.

A \emph{Sequential Model Checker} is a tool which receives a sequential program as input, and checks whether the program is safe. If it is, it returns ``SAFE''. Otherwise, it returns a \emph{counterexample} in the form of a violating path.

\section{Our Methodology}
In this section we describe our methodology for verifying safety properties of concurrent programs, given via assertions.
The main idea is to use a sequential model checker in order to verify the concurrent program.
Our approach handles any (fixed) number of threads. However, for simplicity, we describe our approach for a concurrent program with two threads.
The extension to any number of threads can be found
\ifnoappendix
in~\cite{DannyMSc}.
\else
in \Cref{sec:multiple}.
\fi

In the sequel, we fix a concurrent program $P$ with two threads. We refer to one as \emph{the main thread} (\TMAIN{}) and to the other as \emph{the environment thread} (\TENV{}), with variables $\VMAIN$ and $\VENV$ and program location variables $\PCMAIN$ and $\PCENV$, respectively. $\VMAIN$ and $\VENV$ might intersect. Let $\VORIG = \VMAIN \cup \VENV $. Given a state $s=(\overline{l},\val)$, we denote by $l_M(s)$ and $l_E(s)$ the values of $\PCMAIN$ and $\PCENV$, respectively.
For simplicity, we assume that safety of $P$ is specified by assertions in \TMAIN{} (this is not a real restriction of our method).

Our algorithm generates and maintains a sequential program for each thread. Let $\PAMAIN$ and $\PAENV$ be the two sequential programs, with variables $\VAMAIN \supseteq \VMAIN$ and $\VAENV \supseteq \VENV$. 
Each sequential program might include variables of the other thread as well, together with additional auxiliary variables not in $\VORIG$. Our approach is asymmetric, meaning that $\PAMAIN$ and $\PAENV$ have different roles in the algorithm. $\PAMAIN$ is based on the code of \TMAIN{}, and uses a designated function, \ENVMOVE{}, to abstract computations of \TENV{}. $\PAENV$ is based on the code of \TENV{}, and is constructed in order to answer specific queries for information required by $\PAMAIN$, specified via assumptions and assertions. The algorithm iteratively applies model checking to each of these programs separately. In each iteration, the code of $\PAMAIN$ is gradually modified, as the algorithm learns new information about the environment, and the code of $\PAENV$ is adapted to answer the query of interest.

In \Cref{sec:analyzing_main}, we first describe the way our algorithm operates on $\PAMAIN$.
During the analysis of $\PAMAIN$, information about the environment is retrieved using \emph{environment queries}: 
Intuitively, an environment query receives two conditions, $\alpha$ and $\beta$, and checks whether there exists a reachable computation of \TENV{} in $P$ from $\alpha$ to $\beta$. The idea is to perform specific guided queries in \TENV{}, to search for computations that might ``help'' \TMAIN{} to reach a violation. If such a computation exists, the environment query returns a formula $\psi$, which ensures that all states satisfying it can reach $\beta$ using \TENV{} only. We also require that $\alpha$ and $\psi$ overlap. In order to ensure the reachability of $\beta$, the formula $\psi$ might need to address local variables of \TENV{}, as well as $\PCENV$. These variables will then be added to $\PAMAIN$, and may be used for the input of future environment queries.
If no such computation of the environment exists, the environment query returns $\psi = \FALSE$.
\Cref{subsec:env_queries} describes how our algorithm answers environment queries.
The formal definition follows.

\begin{defn}[Environment Query]\label{defn:env_query}
An \emph{environment query} $\reachenv(\alpha,\beta)$ receives conditions $\alpha$ and $\beta$ over $\VORIGANDPC$, and returns a formula $\psi$ over $\VORIGANDPC$ such that: 
\begin{enumerate}
    \item If there exists a computation of \TENV{} in $P$ that is (1) reachable in $P$, (2) starts from a state $s$ s.t. $s \vDash \alpha$ and (3) ends in a state $s'$ s.t. $s' \vDash \beta$, then $\psi \land \alpha \not \equiv \FALSE$.
    \item If $\psi \not \equiv \FALSE$ then $\alpha \land \psi \not \equiv \FALSE$ and for every state $s$ s.t. $s \vDash \psi$, there exists a computation (not necessarily reachable) of \TENV{} in $P$ from $s$ to some $s'$ s.t. $s' \vDash \beta$.
\end{enumerate}
\end{defn}

\paragraph{Multiple threads.} The key ingredients used by our technique are
\begin{inparaenum}[(i)]
\item an \ENVMOVE{} function that is used in $\PAMAIN$ to overapproximate finite computations (of any length) of \TENV{} (see \Cref{sec:analyzing_main}), and
\item a \TSTART{} function that is used in $\PAENV$ to overapproximate initial computations of $P$ in order to let $\PAENV$ simulate non-initial computations of \TENV{} that follow them (see \Cref{subsec:env_queries}).
\end{inparaenum}
When $P$ has more than two threads, the environment of \TMAIN{} consists of multiple threads, hence environment queries are evaluated by a recursive application of the same approach. Since the computations we consider in the environment are not necessarily initial, the main thread of the environment should now include both the \ENVMOVE{} function and the \TSTART{} function. For more details
\ifnoappendix
see~\cite{DannyMSc}.
\else
see \Cref{sec:multiple}.
\fi

\section{Analyzing the Main thread}\label{sec:analyzing_main}
In this section we describe our algorithm for analyzing the main thread of $P$ for the purpose of proving \longversion{the concurrent program }$P$ safe or unsafe (\Cref{maincode}). \Cref{maincode} maintains a sequential program, $\PAMAIN$, over $\VAMAIN \supseteq \VMAIN$, which represents the composition of \TMAIN{} with an abstraction of \TENV{}. The algorithm changes the code of $\PAMAIN$ iteratively, by adding new assumptions and assertions, as it learns new information about the environment.

\begin{algorithm}[t]
\caption{Algorithm MainThreadCheck}\label{maincode}
\begin{algorithmic}[1]
\Procedure{MainThreadCheck}{\TMAIN{}, \TENV{}, $\phi_{init}$}
\State $\PAMAIN$ = add \ENVMOVE{} calls in \TMAIN{} and initialize \ENVMOVE{}()
\While {a violating path exists in $\PAMAIN$}   \qquad // using sequential MC \label{mainalg:line_while}
    \State Let $\pi=\seqcomma{l}{0}{n+1}$ be a path violating \scode{assert($b$)}.
    \If {there are no \ENVMOVE{}s in $\pi$} {\Return ``Real Violation''}  \label{mainalg:line_has_envmoves}
                                     \label{mainalg:line_real_error}
    \EndIf
    \State let $l_k$ be the label of the last \ENVMOVE{} call in $\pi$
    \State let $\pi_{start} = \seqcomma{l}{0}{k}$ and  $\pi_{end} = \seqcomma{l}{k+1}{n}$
    \State $\beta = wp(\pi_{end}, \neg b)$ \ \ \ \ \qquad \qquad // see (1) in \Cref{subsec:path-analysis}              \label{mainalg:line_beta}
    \State $\alpha = post(\pi_{start}, \phi_{init})$ \ \ \ \qquad // see (2) in \Cref{subsec:path-analysis}   \label{mainalg:line_alpha}
    \State Let $\psi = \reachenv(\alpha,\beta)$  \qquad // environment query for \TENV{} (see \Cref{subsec:env_queries})     \label{mainalg:line_envquery}
    \If {$\psi$ is $\FALSE$}                                                                    \label{mainalg:line_check_psi}
        \State Let $(\alpha', \beta') = \gen(\alpha,\beta)$ \qquad \qquad \qquad \quad //  see (4) in \Cref{subsec:path-analysis}.            \label{mainalg:line_generalize}
        \State $\PAMAIN$ = \scode{RefineEnvMove($\PAMAIN, \alpha', \beta'$)} \qquad \ // see (4) in \Cref{subsec:path-analysis}         \label{mainalg:line_refine}
    \Else{}    \qquad // see (5) in \Cref{subsec:path-analysis}
        \State Add \scode{assert($\neg \psi$)} in $\PAMAIN$ at new label $l'$ right before $l_k$                   \label{mainalg:line_promise_of_violation}
    \EndIf
\EndWhile
\State \Return ``Program is Safe''.   \label{mainalg:line_safe}
\EndProcedure
\end{algorithmic}
\end{algorithm}

The abstraction of \TENV{} is achieved by introducing a new function, \ENVMOVE{}. Context switches from \TMAIN{} to \TENV{} are modeled explicitly by calls to \ENVMOVE{}. The body of \ENVMOVE{} changes during the run of \Cref{maincode}. However, it always has the property that it over-approximates the set of finite (possibly of length zero) computations of \TENV{} in $P$ that are reachable in $P$. This is formalized as follows:

\begin{defn}[Overapproximation]\label{lem:env_move_over}
For a state $s_m$ of $\PAMAIN$ (over $\VAMAIN$) s.t. $l(s_m)$ is the beginning or the end of \ENVMOVE{}, we say that $s_m$ \emph{matches} a state $s$ of $P$ (over $V$) if
\begin{inparaenum}[(1)]
\item $s_m$ and $s$  agree on $\VAMAIN \cap \VORIG$, i.e. $\val(s_m)|_{\VORIG} = \val(s)|_{\VAMAIN}$,
        where $\val|_{U}$ is the projection of $\val$ to the variables appearing in $U$, and
\item if $\PCENV \in \VAMAIN$, then $\val(s_m)(\PCENV) = l_E(s)$.
\end{inparaenum}

We say that \emph{\ENVMOVE{} overapproximates the computations of \TENV{} in $P$} if for every  reachable computation $\COMP=s\xrightarrow{t_{E}}\dots\xrightarrow{t_{E}}s'$ of \TENV{} in $P$ (possibly of length $0$), and for every state $s_m$ s.t. $l(s_m)$ is the beginning of \ENVMOVE{} and $s_m$ matches $s$,
there exists a
computation $\COMP_m=s_m\rightarrow\dots\rightarrow s_m'$ of $\PAMAIN$ s.t.
\begin{inparaenum}[(1)]
\item $\COMP_m$ is a complete execution of \ENVMOVE{}, i.e., 
$l(s_m')$ is the end of \ENVMOVE{} and for every other state $s_m''$ in $\COMP_m$, $l(s_m'')$ is a label within \ENVMOVE{}, and
\item $s_m'$ matches $s'$.
\end{inparaenum}
\end{defn}

The code of $\PAMAIN$ always consists of the original code of \TMAIN{}, the body of the \ENVMOVE{} function (which contains assumptions about the environment), calls to \ENVMOVE{} that are added at initialization, and new assertions that are added during the algorithm. $\VAMAIN$ always consists of $\VMAIN$, possibly $\PCENV$, some variables of $\VENV$ (that are gradually added by need), and some additional auxiliary variables needed for the algorithm (see (4) in \Cref{subsec:path-analysis}).

\vspace{-0.3cm}
\subsection{Initialization}
\vspace{-0.1cm}
\Cref{maincode} starts by constructing the initial version of $\PAMAIN$, based on the code of \TMAIN{}. To do so, it adds explicit calls to \ENVMOVE{} at every location where a context switch needs to be considered in \TMAIN{}.
The latter set of locations is determined by an \emph{interleaving reduction analysis}, which
identifies a set of locations, called \emph{cut-points}, such that the original program is safe if and only if all the computations in which context-switches occur only at cut-points are safe.

In addition, the algorithm constructs the initial \ENVMOVE{} function which havocs 
every shared variable of \TENV{} and \TMAIN{} that is written by \TENV{}. 
This function will gradually be refined to represent the environment in a more precise way.

\lstset{ %
  escapechar=@,
  numbers=left,                    
  numberstyle=\tiny,
  basicstyle=\scriptsize
}

\begin{figure}[t]
\begin{minipage}{\textwidth}
\lstset{firstnumber=last}
\begin{tabular}{l}
\begin{lstlisting}[language=C++] %, frame=single]

bool claim0 = false, claim1 = false;
bool cs1 = false, cs0 = false;
int turn;
\end{lstlisting}
\end{tabular}\\
\begin{tabular}{ll}
\begin{lstlisting}[language=C++]
void t0() {  @\label{example:t0_proc}@
  while (true) {                          @\label{example:t0_main_loop}@
    claim0 = true;                        @\label{example:t0_claim}@
    turn = 1;                             @\label{example:t0_pass_turn}@
    while (claim1 && turn != 0) { }        @\label{example:t0_busy_wait}@
    cs0 = true;                            @\label{example:t0_c0_true}@
    // CRITICAL_SECTION
    @\hl{\texttt{assert(!cs1)} ;}@         @\label{example:t0_env_move_assert}@
    cs0 = false;                           @\label{example:t0_c0_false}@
    claim0 = false;     }}          @\label{example:t0_unclaim}@
\end{lstlisting}
&
\begin{lstlisting}[language=C++] %,firstnumber=last]

void t1() {
  while (true) {                            @\label{example:t1_main_loop}@
    claim1 = true;
    turn = 0;                               @\label{example:t1_pass_turn}@
    while (claim0 && turn != 1) { }         @\label{example:t1_busy_wait}@
    cs1 = true;                             @\label{example:t1_cs1_true}@
    // CRITICAL_SECTION
    cs1 = false;                            @\label{example:t1_cs1_false}@
    claim1 = false;      }}       @\label{example:t1_unclaim} \label{example:t1_main_loop_leave}@
\end{lstlisting}
\end{tabular}
\end{minipage}
\caption{Peterson's mutual exclusion algorithm for two threads $t0$ and $t1$.
}\label{fig:Peterson}
\end{figure}

\begin{exmpl} \label[exmpl]{ex:peterson}
We use Peterson's algorithm \cite{DBLP:journals/ipl/Peterson81} for mutual exclusion, presented in \Cref{fig:Peterson}, as a running example. The algorithm contains a busy-wait loop in both threads, where a thread leaves that loop and enters its critical section only after the \scode{turn} variable indicates that it is its turn to enter, or the other thread gave up on its claim to enter the critical section. In order to specify the safety property (mutual exclusion), we use additional variables \scode{cs0}, \scode{cs1} which indicate that \scode{t0} and \scode{t1} (resp.) are in their critical sections.
The safety property is that $\neg cs0 \vee \neg cs1$ always holds.
It is specified by the \scode{assert(!cs1)} command in $t0$ between lines \ref{example:t0_c0_true} and \ref{example:t0_c0_false}, where \scode{cs0} is \scode{true}\longversion{\footnote{This is sufficient, as $cs0$ is clearly $\FALSE$ at every other location}.}

Assume that $t0$ was chosen as the main thread and $t1$ as the environment thread. We generate a sequential program $P_0$, based on the code of $t0$: we add \ENVMOVE{}s at every cut point, as determined by our interleaving reduction mechanism.
\longversion{In our case, the cut points are after all commands except for \ref{example:t0_c0_true} and \ref{example:t0_c0_false} 
which change a local variable of $t0$.}%
\longversion{
The result is illustrated by \Cref{fig:add_envmove}. Intuitively, we can see that no \ENVMOVE{} is required after writing to \scode{cs0} in lines \ref{example:t0_env_move_c0_true} and \ref{example:t0_env_move_c0_false}, since it is a local variable of $t0$
\footnote{In fact, since the condition \scode{claim1 \&\& turn != 0} in line \ref{example:t0_env_move_busy_wait} of \Cref{fig:add_envmove} is not evaluated atomically in C programs, another \ENVMOVE{} is required after reading \scode{claim1} and before reading \scode{turn}. We achieve this by rewriting the program, s.t. it first assigns the values of \scode{claim1} and \scode{turn} to two new local variables, calls \ENVMOVE{} between these two assignments, and only evaluates the new local variables to check whether the condition holds. We omit this here for simplicity.}.
}%
The initial \ENVMOVE{} only havocs all variables of $P_0$ that are written by $t1$, i.e., \scode{claim1}, \scode{turn}, \scode{cs1} (see \Cref{fig:env_move_init-refine}). 

\end{exmpl}

\subsection{Iteration of the MainThreadCheck Algorithm} \label{subsec:mainalg}

Each iteration of \Cref{maincode} starts
by applying a sequential model checker to check whether there exists a violating path (that may involve calls to \ENVMOVE{}) in $\PAMAIN$ (line \ref{mainalg:line_while}). If not, we conclude that the \emph{concurrent} program is safe (line \ref{mainalg:line_safe}), as the \ENVMOVE{} function over-approximates the computations of the environment.
If an assertion violation is detected in $\PAMAIN$, the model checker returns a counterexample in the form of a violating path. If there are no \ENVMOVE{} calls in the path (line \ref{mainalg:line_has_envmoves}), it means that the path represents a genuine violation obtained by a computation of the original main thread, and hence the program is unsafe. 

Otherwise, the violation relies on environment moves, and as such it might be spurious. We therefore analyze this counterexample as described in \Cref{subsec:path-analysis}. The purpose of the analysis is to check whether \TENV{} indeed enables the environment transitions used along the path. If so, we find ``promises of error'' for the violated assertion at earlier stages along the path and add them as new assertions in $\PAMAIN$. Intuitively speaking, a ``promise of error'' is a property ensuring that \TENV{} can make a sequence of steps that will allow \TMAIN{} to violate its assertion. Such a property may depend on both threads, and hence it is defined over $\VORIGANDPC$ ($\PCMAIN$ is given implicitly by the location of the assertion in $\PAMAIN$).
Formally, we have the following definition:

\begin{defn} \label{defn:promise}
    Let $\psi, \psi'$ be formulas over $\VORIGANDPC$
    and let $l,l'$ be labels of \TMAIN. We say that $(l,\psi)$ is a \emph{promise} of $(l',\psi')$ if for every state $s$ of $P$ s.t. $l_M(s)=l$ and $s \vDash \psi$ there exists a computation in $P$ starting from $s$ to a state $s'$ s.t. $\l_M(s')=l'$ and $s' \vDash \psi'$.

    If $(l,\psi)$ is a \emph{promise} of $(l',\neg b)$ and $l'$ has an \scode{assert($b$)} command, then we say that $(l,\psi)$ is a \emph{promise of error}.
\end{defn}
\longversion{A promise (and a promise of error) $(l,\psi)$ is defined ``from the point of view'' of \TMAIN{}, i.e., if $\psi$ holds when \TMAIN{} is at a specific location $l$, then $(l',\psi')$ (or an error) can be reached. Note that the definition refers to computations of $P$, and is independent of our construction of $\PAMAIN$ and $\PAENV$.   
}
Note \longversion{further }that the definition is transitive. Specifically, if $(l,\psi)$ is a \emph{promise} of $(l',\psi')$ and $(l',\psi')$ is a promise of error, then $(l,\psi)$ is also a promise of error.

\paragraph{Outcome.} Each iteration of \Cref{maincode} ends with one of these three scenarios:
\begin{compactenum}
  \item The algorithm terminates having found a genuine counterexample for $P$ (line \ref{mainalg:line_real_error}). 
  \item The obtained counterexample is found to be spurious since an execution of \ENVMOVE{} along the path is proved to be infeasible. The counterexample is eliminated by refining the \ENVMOVE{} function (line \ref{mainalg:line_refine}, also see item $(4)$ in the next section).
  \item Spuriousness of the counterexample remains undetermined, but a new promise of error is generated before the last \ENVMOVE{} call in the violating path\longversion{ of $\PAMAIN$}. We augment $\PAMAIN$ with a new assertion, representing this promise of error (line \ref{mainalg:line_promise_of_violation}).
\end{compactenum}

\longversion{
\paragraph{}
We explain in detail the analysis of a potentially spurious violating path of $\PAMAIN$, as well as the generation of new promises of error, in \Cref{subsec:path-analysis} and the refinement of \ENVMOVE{} when the path is spurious  in \Cref{subsec:envmove}.
}

\subsection{Analyzing a potentially spurious violating path} \label{subsec:path-analysis}

\longversion{This subsection thoroughly explains 
the analysis of a potentially spurious violating path in $\PAMAIN$, i.e., a path that contains at least one call to \ENVMOVE{},  obtained in an iteration of \Cref{maincode}.
}
Let $\pi=\seqcomma{l}{0}{n+1}$ be a violating path of $\PAMAIN$,
returned by the sequential model checker in an iteration of \Cref{maincode},
which is potentially spurious in $P$, i.e., 
contains at least one \ENVMOVE{} call. 
Since $\pi$ 
is violating, $l_{n+1}=l_\epsilon$ and $\act(l_n)=$\scode{assert($b$)} for some condition $b$.
Let $l_k$, for some $0 \leq k \leq (n-1)$, be the location of the last \ENVMOVE{} in 
$\pi$. We perform the following steps, illustrated by \Cref{fig:alpha-beta}:

\begin{figure}[t]
\begin{tabular}{cc}
\includegraphics[width=0.5\textwidth]{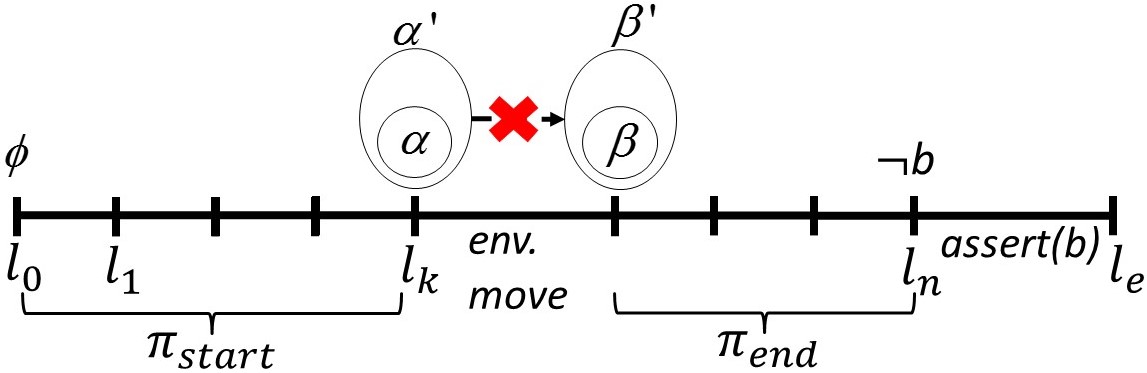} \hspace{0.3cm}
&
\includegraphics[width=0.5\textwidth]{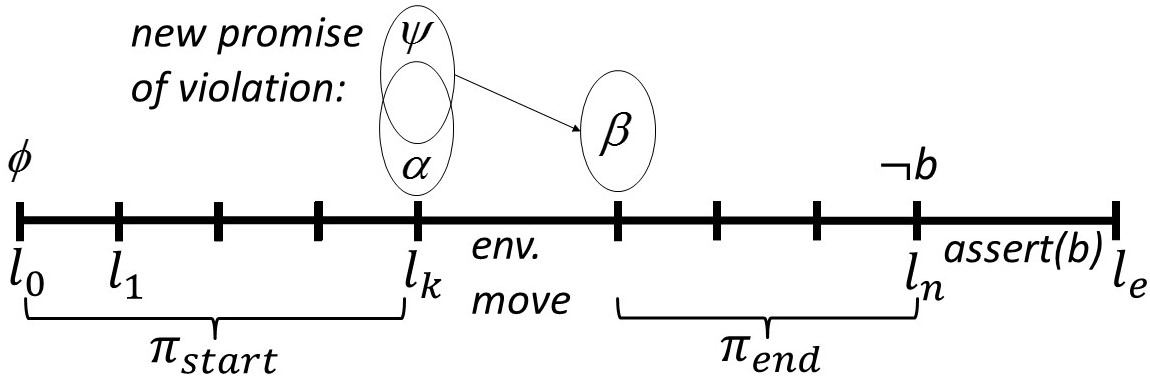}
\\ (a) & (b)
\end{tabular}
\caption{(a) If $\reachenv(\alpha,\beta)=\FALSE$, we search for more general $\alpha'$ and $\beta'$ which restrict the environment transition; (b) If $\reachenv(\alpha,\beta)=\psi \neq \FALSE$, then we know that $\psi$ leads to $\beta$ and that $\psi \land \alpha \neq \FALSE$.\label{fig:alpha-beta}}
\end{figure}

\lstset{ %
  escapechar=@,
  numbers=left,                    
  numberstyle=\tiny,
  basicstyle=\scriptsize
}

\begin{figure}[t]
\begin{center}
\begin{minipage}{.4\textwidth}
\begin{lstlisting}[language=C++, frame=single,breaklines = {true}]
void P0() {
  assert((!cs1) || claim1); @\label{example_part:assert_first}@
  env_move(); @\label{example_part:t0_env_move_before_main_loop}@
  while (true) { @\label{example_part:t0_env_move_main_loop}@
    claim0 = true; @\label{example_part:t0_env_move_claim}@
    assert((!cs1) || claim1); @\label{example_part:assert_after_claim}@
    env_move();  @\label{example_part:t0_env_move_before_pass_turn}@
    turn = 1;  @\label{example_part:t0_env_move_pass_turn}@
    assert((!cs1) ||             (claim1 && turn != 0)); @\label{example_part:last_assert}@
    ...
  }}
\end{lstlisting}
\caption{The sequential program $P_0$ after a few iterations of \Cref{maincode}.}\label{fig:main_partial}
\end{minipage}
\hspace{1cm}
\begin{minipage}{.4\textwidth}
\begin{lstlisting}[language=C++, frame=single]
void env_move() {
  bool claim1_copy = claim1;
  int turn_copy = turn;
  bool cs1_copy = cs1;
  claim1 = havoc_bool();
  turn = havoc_int();
  cs1 = havoc_bool();
  @\hl{if (true) \{ assume(!cs1 $||$ claim1) ; \} }@ }
\end{lstlisting}
\caption{The \ENVMOVE{} function of $P_0$: initially (without highlighted lines); and after one refinement (with highlighted lines).}\label{fig:env_move_init-refine}
\end{minipage}
\end{center}
\end{figure}

\paragraph{(1) Computing condition after the environment step:} We compute (backwards) the weakest precondition 
of $\neg b$ w.r.t. the path
  $\pi_{end}=\seqcomma{l}{k+1}{n}$ 
  to obtain
  $\beta=wp(\pi_{end}, \neg b)$ (line \ref{mainalg:line_beta}).
  Recall that $\neg b$ is necessarily reachable from $\beta$ along $\pi_{end}$ in $\PAMAIN$.%
  \longversion{
	  Recall that $\beta$ has the property that for every state $s$, $\val(s) \vDash \beta$ iff there exists a computation in $\PAMAIN$ starting from $s$ that passes through the path $\pi_{end}$ and reaches a state $s'$ s.t $\val(s') \vDash \neg b$.
	} \label{alg_outline:beta}

\paragraph{(2) Computing condition before the environment step:}
  We compute (forward) a postcondition $\alpha = post(\pi_{start}, \phi_{init}) $ starting from $\phi_{init}$ for the path $\pi_{start}=\seqcomma{l}{0}{k}$ (line  \ref{mainalg:line_alpha}).
  To ensure progress, we make sure that if $\pi_{start}$ ends with a suffix of \scode{assert}s then $\alpha \implies c$ for every {\scode{assert($c$)} command that appears in this suffix (e.g., by conjoining $\alpha$ with $c$).
  Recall that $\alpha$ necessarily holds after executing $\pi_{start}$ in $\PAMAIN$ from $\phi_{init}$.
\longversion{
Recall that $\alpha$ has the property that for every computation from state $s$ to state $s'$ whose path is $\seqcomma{l}{0}{k-1}$, if $\val(s) \vDash \phi_{init}$ then $\val(s') \vDash \alpha$.
}
  \label{alg_outline:alpha}

\paragraph{(3) Environment query: }We compute $\psi = \reachenv(\alpha,\beta)$ (line \ref{mainalg:line_envquery}).
    \label{alg_outline:env_query}

    \begin{exmpl} 
\label[exmpl]{example:alg_false_env}

\Cref{fig:main_partial} presents a prefix of $\PAMAIN$ after a few iterations of the algorithm, before the first refinement of \ENVMOVE{} (i.e., $\PAMAIN$ still uses the initial \ENVMOVE{} function). The previous iterations found new promises of error, and augmented $\PAMAIN$ with new assertions. Consider the initial conditions from \Cref{fig:Peterson}, i.e., $\phi_{init}\triangleq [\scode{claim0 = }$ $\scode{claim1  = cs1 = cs0 = false}]$.
Assume that our sequential model checker found the violation given by the next path:
\ref{example_part:assert_first},
\ref{example_part:t0_env_move_before_main_loop},
\ref{example_part:t0_env_move_main_loop},
\ref{example_part:t0_env_move_claim},
\ref{example_part:assert_after_claim},
\ref{example_part:t0_env_move_before_pass_turn},
\ref{example_part:t0_env_move_pass_turn},
\ref{example_part:last_assert},
\longversion{reaching and violating \scode{assert(!cs1 || (claim1 \&\& turn != 0))}}\longversion{ in line \ref{example_part:last_assert}}.

To check whether the last \ENVMOVE{} call in line \ref{example_part:t0_env_move_before_pass_turn} represents a real computation of \scode{t1}, we compute
the weakest precondition of the condition $\neg b \triangleq \scode{cs1} \land (\neg \scode{claim1} \lor \scode{turn}=0)$, taken from the violated assertion in line \ref{example_part:last_assert}, w.r.t. the path $\pi_{end}=\ref{example_part:t0_env_move_pass_turn},
\ref{example_part:last_assert}$.
The result is $\beta = wp(\pi_{end}, \neg b) = (\scode{cs1} \land \neg \scode{claim1})$.
The computation of $\alpha=post(\pi_{start},\phi)$ for the path $\pi_{start}=\ref{example_part:assert_first},
\ref{example_part:t0_env_move_before_main_loop},
\ref{example_part:t0_env_move_main_loop},
\ref{example_part:t0_env_move_claim},
\ref{example_part:assert_after_claim}$ yields $\alpha = (\neg \scode{cs0} \land \scode{claim0} \land  (\neg \scode{cs1} \lor \scode{claim1}))$.
We then generate an environment query $\reachenv(\alpha,\beta)$.
\end{exmpl}

\paragraph{(4) Refining the \ENVMOVE{} function: } If $\psi = \FALSE$ (line \ref{mainalg:line_check_psi}) it means that there is no reachable computation of \TENV{} in $P$ from a state $s$ s.t. $s \models \alpha$ to a state $s'$ s.t. $s' \models \beta$.
We apply a generalization procedure $\gen(\alpha,\beta)$ that returns 
$\alpha',\beta'$ s.t. $\alpha \implies \alpha', \beta \implies \beta'$ and still $\reachenv(\alpha',\beta') = \FALSE$ (line \ref{mainalg:line_generalize}).
To do so, $\gen$ 
iteratively replaces $\alpha$ and/or $\beta$ with $\alpha',\beta'$ s.t.  $\alpha \implies \alpha', \beta \implies \beta'$ and rechecks $\reachenv(\alpha',\beta')$.
For example, if $\alpha$ contains a subformula of the form $\delta_1 \wedge \delta_2$ that appears positively, we attempt to replace it by $\delta_1$ or $\delta_2$ to obtain $\alpha'$.\footnote{More information about the generalization appears in the optimizations section in \cite{DannyMSc}.}
We then refine \ENVMOVE{} to eliminate the environment transition from $\alpha'$ to $\beta'$ (line \ref{mainalg:line_refine}).
\Cref{fig:alpha-beta}(a) illustrates this step.

The refinement is done by
introducing in \ENVMOVE{}, after the variables are havocked, the command (\scode{if ($\alpha'(W$\_old)) assume($\neg \beta'$)}), where  $W$\_old are the values of the variables before they are havocked in \ENVMOVE{} (these values are copied by \ENVMOVE{} to allow evaluating $\alpha'$ on the values of the variables before \ENVMOVE{} is called). 
The command blocks all computations of $\ENVMOVE{}$ from $\alpha'$ to $\beta'$. Since such computations were proven by the environment query to be infeasible in \TENV{}, we are ensured
that \ENVMOVE{} remains an overapproximation of the computations of \TENV{}.

\begin{exmpl}\label[exmpl]{ex:refine}
The call to $\reachenv(\alpha,\beta)$ in \Cref{example:alg_false_env} results in $\psi=\FALSE$. Hence, we apply generalization. We obtain two formulas $\alpha'=TRUE,\beta'=\beta$ which indeed satisfy  $\alpha \implies \alpha'$, $\beta \implies \beta'$
and $\reachenv(\alpha',\beta')=\FALSE$.
This means that when \TENV{} is called with $\alpha' = TRUE$, then no computation of \TENV{} reaches a state satisfying $\beta' = cs1 \land \neg claim1$.
\Cref{fig:env_move_init-refine} presents the \ENVMOVE{} function before and after the refinement step based on $(\alpha',\beta')$ takes place.
The refinement step
adds the highlighted line to the initial \ENVMOVE{} function.
This line has the constraint 
\scode{if (true) assume(!cs1 $||$ claim1)}, derived from the observation above.

\end{exmpl}

     \label{alg_outline:nopath}

\paragraph{(5) Adding assertions:} If $\psi \neq \FALSE$, then for every state satisfying $\psi$ there is a computation of \TENV{} in $P$ to a state satisfying $\beta$. Since $\beta=wp(\pi_{end}, \neg b)$, it is guaranteed that this computation can be extended (in \TMAIN{}) along the path $\pi_{end}$, which does not use any environment moves, to reach a state $s'$ that violates the assertion \scode{assert(b)}.
      This is illustrated in \Cref{fig:alpha-beta}(b).
      We therefore conclude that if $\psi$ is satisfied before the \ENVMOVE{} at label $l_k$, a genuine violation can be reached, making $(\hat{l_k},\psi)$ a promise of error, where $\hat{l_k}$ denotes the label in \TMAIN{}
       that corresponds to $l_k$ (the label reached after executing the \ENVMOVE{} called at label $l_k$).
      Therefore, we add a new assertion \scode{assert($\neg \psi$)} 
      right before $l_k$
      (line \ref{mainalg:line_promise_of_violation}).
      In addition, if $\psi$ includes a variable $v$ that is not in $\VAMAIN$ (e.g., $\PCENV$), then $v$ is added to $\VAMAIN$, its declaration (and initialization, if exists) is added to $\PAMAIN$, and  \ENVMOVE{} is extended to havoc $v$ as well (if it is written by \TENV{}).
      \label{alg_outline:path_exist}

\section{Answering Environment Queries} \label{subsec:env_queries}
Recall that an environment query $\reachenv(\alpha,\beta)$ checks whether there exists a reachable computation $\COMP$ of \TENV{} in $P$ from a state $s \models \alpha$ to a state $s' \models \beta$. This computation may involve any finite number of steps of \TENV{}, executed without interference of \TMAIN{}.

If $\alpha \land \beta \not \equiv \FALSE$, we simply return $\beta$, which represents a computation of length zero.
Otherwise, we wish to apply a sequential model checker on \TENV{} in order to reveal such computations, or conclude there are none. However,
the computation $\COMP$ may not be initial,
while our sequential model checker can only search for violating paths starting from an initial state. Hence we construct a modified sequential program $\PAENV$, based on the code of \TENV{}, which also represents (over-approximates) non-initial, but reachable, computations $\COMP$ of \TENV{} in $P$.
For that, we add in $\PAENV$ calls to a new function, \TSTART{}, which models the runs of \TMAIN{} until the start of $\COMP$.
The calls to \TSTART{} are added in all cut-points computed by an interleaving reduction (similar to the one applied to~\TMAIN{}).

\paragraph{The \TSTART{} function.}
The \TSTART{} function is responsible for non-deterministically setting the start point of $\COMP$, where context switches to \TMAIN{} are no longer allowed.
This is done by setting a new \scode{start} variable to true (provided that its value is not yet true).
We refer to the latter call as the \textit{activation} \TSTART{}.
As long as \scode{start} is false (i.e., prior to the activation call), \TSTART{} havocs the variables written by \TMAIN{}. When \scode{start} is set to true, we add an \scode{assume($\alpha$)} command after the havoc commands as this is the state chosen to start the computation.
To handle the case where $\PCENV$ appears in $\alpha$, \TSTART{} receives the original location (in \TENV{}) in which it is called as a parameter, and updates the explicit $\PCENV$ variable. Whenever \scode{start} is already true, \TSTART{} immediately exits,
ensuring that $\COMP$ indeed only uses transitions of~\TENV{}.

In $\PAENV$, we also add assertions of the form \scode{assert(!start || $\neg \beta$)} after every call to \TSTART{}. Hence, a violating path, if found, reaches $\scode{start}\land\beta$, i.e., it captures a computation in which $\alpha$ was satisfied (when \scode{start} was set to true), and reached $\beta$.

\paragraph{Returning Result.}
If a violating path is not found, we return $\reachenv(\alpha,\beta)=\FALSE$. If a violating path $\seqcomma{m}{0}{n+1}$ is found, let $m_k$ be the label of the activation \TSTART{} for some $0 \leq k \leq (n-1)$.
Let $\pi_E$ be the projection of $\seqcomma{m}{k+1}{n-1}$ to \longversion{the commands from }\TENV{}.
We compute the weakest precondition of $\beta$ w.r.t. the path $\pi_E$ and obtain $\psi = wp(\pi_{E}, \beta)$. The computed $\psi$ satisfies the desired requirement: For every state $s$ of $P$ s.t. $s \vDash wp(\pi_{E}, \beta)$, there exists a computation $\COMP$ of \TENV{} starting from $s$ which follows the path $\pi_{E}$ and reaches a state $s'$ satisfying $\beta$.
%
Note that $\COMP$ might not be reachable, as in the prefix we used an abstraction of \TMAIN{}. That means that $\reachenv(\alpha,\beta)$ is not ``exact'' and may return $\psi \neq \FALSE$ when there is no reachable computations as required. However, it satisfies the requirements of \Cref{defn:env_query}, which is sufficient for soundness and progress.
The intuition is that checking the reachability of $\psi$ is done by the main thread.

For an example demonstrating how an environment query is answered see \cite{DannyMSc}.

\section{Soundness and Progress} \label{subsec:termination}
Our algorithm for verifying the concurrent program $P$ terminates when either
\begin{inparaenum}[(i)]
  \item all the assertions in $\PAMAIN$ are proven safe (i.e., neither the original error nor all the new promises of error can be reached in $\PAMAIN$), in which case
  \Cref{maincode} returns ``Program is Safe''.
  \item a violation of some assertion in $\PAMAIN$, which indicates either the original error or a promise of error, is reached without any \ENVMOVE{} calls, in which case
  \Cref{maincode} returns ``Real Violation''.
\end{inparaenum}
The following theorem summarizes its soundness\footnote{Full proofs appear in \href{https://tinyurl.com/comusfull}{https://tinyurl.com/comusfull}.}.
\begin{theorem}
If \Cref{maincode} returns ``Safe'' then the concurrent program $P$ has no violating computation; If it returns ``Real violation'' then $P$ has a violating computation.
\end{theorem}
The proof of the first claim shows that our algorithm maintains the overapproximation property of \ENVMOVE{} (see \Cref{lem:env_move_over}), from which the claim follows immediately. In the proof of the second claim, we show that the properties of an environment query (see \Cref{defn:env_query}) and of promises of errors (\Cref{defn:promise}) are satisfied.

\paragraph{}
While termination is not guaranteed for programs over infinite domains, the algorithm is ensured to make \emph{progress} in the following sense.
Each iteration either refines \ENVMOVE{} (step $(4)$ in \Cref{subsec:path-analysis}), making it more precise w.r.t. the real environment, or generates new promises of errors at earlier stages along the violating path (step $(5)$ in \Cref{subsec:path-analysis}).
In the former case, the set of pairs of states $(s,s')$ represented by the start and end states of computations of \ENVMOVE{} is strictly decreasing -- this set overapproximates the set of pairs of states $(s, s')$ for which \TENV{} has a reachable computation from state $s$ to state $s'$ (see \Cref{lem:env_move_over}). 
In the latter case, the set of states known to lead to a real violation of safety is strictly increasing. In both cases, the other set remains unchanged.

When the domain of all variables is finite, these two sets are bounded, hence the algorithm is guaranteed to terminate.
\longversion{
Note that the same violating path may recur in the algorithm, however, the set of computations that can be observed along it decreases either since a new assertion (promise of error) was added, or because the \ENVMOVE{} function was refined.
}

\section{Experimental Results and Conclusion}
\label{sec:experimental}

\paragraph{Setup.}
We implemented our algorithm in a prototype tool called CoMuS.
The implementation is written in Python 3.5, uses pycparser \cite{pycparser} for parsing and transforming C programs,
uses SeaHorn~\cite{DBLP:conf/cav/GurfinkelKKN15} for sequential model checking, and uses Z3~\cite{de2008z3} to check logical implications 
for some optimizations.
A description of the optimizations 
can be found
\ifnoappendix
in~\cite{DannyMSc}.
\else
in \Cref{sec:optimizations}.
\fi
CoMuS currently supports only a subset of the syntax of C (see \Cref{sec:prelim}). 
It does not perform alias analysis and hence has limited pointers support.
It also does not support dynamic thread creations, although we support any fixed number of threads.

We compare 
CoMuS with Threader~\cite{DBLP:conf/tacas/PopeeaR13},
VVT~\cite{DBLP:conf/tacas/GuntherLW16} and UL-CSeq~\cite{DBLP:conf/tacas/Nguyen0TP15},
the last two being the top scoring model checkers on the concurrency benchmark among \emph{sound unbounded} tools in SVCOMP'16 and SVCOMP'17 (resp.).
On the concurrency benchmark, VVT was $4^\text{th}$ overall in SVCOMP'16, and UL-CSeq was $8^\text{th}$ overall in SVCOMP'17\ \footnote{The same benchmark was used for unbounded sound tools and tools which perform unsound bounded reductions.
Bounded tools are typically ranked higher. Our method is unbounded and is able to provide proofs,
hence we find the selected tools more suitable for comparison.}.
Threader performs modular verification, abstracts each thread separately and uses an interference abstraction for each pair of threads. UL-CSeq performs a reduction to a single non-deterministic sequential program. We used it in its default mode, with CPAChecker~\cite{DBLP:conf/cav/BeyerK11}
as a backend. VVT combines bounded model checking for bug finding with an IC3 \cite{DBLP:conf/vmcai/Bradley11} based method for full verification.

We ran the experiments on a x86-64 Linux machine, running Ubuntu 16.04 (Xenial) using
Intel(R) Xeon(R) CPU E5-2680 v3 @ 2.50GHz with 8GB of RAM.

\paragraph{Experiments.} 
We evaluated the tools using three experiments. One compares the four tools on concurrent programs with a clear hierarchy. The second compares syntactically similar programs with and without hierarchal structure to evaluate the effect of the structure on the verification time. The last one looked at general concurrent programs.

\paragraph{Hierarchically structured programs.}
For the first experiment, we used three concurrent dynamic-programming algorithms:
Sum-Matrix, Pascal-Triangle and Longest-Increasing-Subsequence.
The Sum-Matrix programs receive a matrix $A$ as input. For every pair of indexes $(i,j)$, it computes the sum of all elements $A[k,l]$, where $k \geq i$ and $l \geq j$.
In their concurrent version, each thread is responsible for the computation of a single row. The Pascal-Triangle programs compute all the binomial coefficients up to a given bound. Each thread computes one row of the triangle, where each element in the row depends on two results of the previous row. The Longest-Increasing-Subsequence programs receive an array, and compute for each index $i$, the length of the longest increasing subsequence that ends at index $i$. Each thread is responsible for computing the result for a given index of the array, depending on the result of all prefixes. Both these and the matrix programs are infinite state, as the elements of the array (resp. the matrix) are unbounded inputs.

These algorithms have a natural definition for any finite number of threads. Typically, the verification becomes harder as the number of threads increases. For evaluation, we used programs with an increasing number of threads, and check the influence of the number on the different tools. For each instance, we use both a safe and an unsafe version. Both versions differ from each other either only by a change of specification, or by a slight modification that introduces a bug.

The chosen programs have two meaningful characteristics:
\begin{inparaenum}[(i)] 
	\item They exhibit non-trivial concurrency.
This means that each thread performs a series of computations, and it can advance when the data for each computation is ready, without waiting for the threads it depends on to complete. Consider the Sum-Matrix problem as an example. Assume thread $t_i$ needs to compute the result at some location $(i,j)$, and that each row is computed backwards (from the last cell to the first).
	The computation exploits the results of thread $t_{i+1}$. Thread
	$t_i$ needs to wait for thread $t_{i+1}$ to compute the result for location $(i+1,j)$. However, $t_i$ does not wait for $t_{i+1}$ to terminate, as it can compute the cell $(i,j)$, while $t_{i+1}$ continues to compute $(i+1,j-1)$. 
	\item Their data flow graph has a clear chain structure. That is,
the threads can be ordered in a chain hierarchy, and each thread only requires information computed by its immediate successor.
\end{inparaenum}

\begin{figure}[t]
   \centering
   \includegraphics[width=0.85\textwidth]{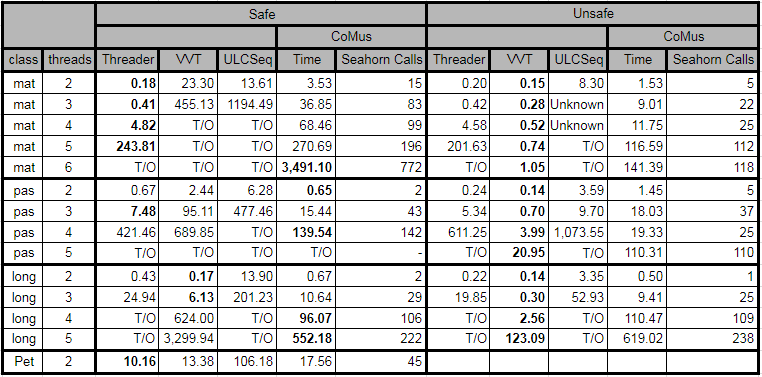}
   \caption{Run times [secs] for all four tools for verifying concurrent dynamic programs algorithms.
   }\label{fig:table}
 \end{figure}

\Cref{fig:table} summarizes the results for these programs.
The timeout 
was set to 3600 seconds. 
The code of the programs 
is available at \url{tinyurl.com/comusatva18}. We include in the table also our running example, the Peterson algorithm.

The results demonstrate a clear advantage for CoMuS for verification (i.e., for safe programs) as the number of threads increases.
This can be attributed to the chain structure that lets CoMuS minimize the amount of information transferred between threads.
For falsification, CoMuS is outperformed by VVT's bounded method. However, it still performs significantly better than the two other tools 
when the number of threads grows.

\paragraph{Hierarchical vs. non-hierarchical programs.}
The programs used for this evaluation are variants of the ``fib\_bench'' examples of the SV-COMP concurrency benchmark.
We compare programs in which the data flow graph has a ring topology, vs. programs in which it has a chain topology.
For the ring case, consider a program with threads $\seqcomma{t}{0}{n-1}$ and variables $\seqcomma{v}{0}{n-1}$. Each thread $t_i$ runs in a loop, and iteratively performs $v_i$\scode{+=}$v_{(i+1 (mod\ n))}$. The checked property is that $v_0$ does not surpass an upper bound. The chain case is identical except that 
for the last thread, $t_{n-1}$, we break the chain and perform $v_{n-1}$\scode{+=}$1$ instead of $v_{n-1}$\scode{+=}$v_{0}$.
\Cref{fig:tablefib} presents the results of this comparison. All the programs in the table are safe and with two loop iterations. The timeout was set to 1200 seconds.

For the ring case, all tools fail to verify programs with $\geq 4$ threads. Threader presents similar results for both ring and chain topologies. VVT benefits from the less dependent chain topology, but still timeouts on 
$> 3$ threads. CoMuS, on the other hand, is designed to exploit hierarchy, and benefits significantly from the chain topology, where it verifies all instances. UL-CSeq is excluded from the table
as it times-out on the ``fib\_bench'' examples
(both in our experiments and in the SV-COMP results).

The reason for CoMus's different runtime on the chain and ring variants is that for programs that have no clear hierarchy (as in the ring programs), the conditions passed to the environment queries must include information relevant to the caller thread; a manual inspection shows that they typically become more complex. As similar phenomenon happens if the verification order used by CoMuS is not aligned with the hierarchy of the program. For example, switching the verification order of the last two threads in the long\_th3\_safe example, increases the verification time from 10 to 25 seconds.

\begin{figure}[t]
	\centering
	\includegraphics[width=0.75\textwidth]{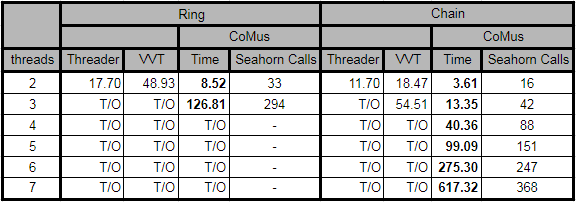}
	\caption{Run times [secs] for fib\_bench programs with ring topology vs. chain topology.
	}\label{fig:tablefib}
\end{figure}

\paragraph{General concurrent programs.}
We also evaluated the tools on a partial subset of the SV-COMP concurrency benchmark, whose code is supported by CoMuS. Typically, on these runs CoMuS was outperformed by the other tools.
We conclude that even though our method can be applied to programs without a clear hierarchical structure, it is particularly beneficial for programs in which the hierarchy is inherent.

\paragraph{Conclusion.}
In this work we develop an automatic, modular and hierarchical method for proving or disproving safety of concurrent programs by exploiting model checking for sequential programs. The method can handle infinite-state programs. It is sound and unbounded.
We implemented our approach in a prototype tool called CoMuS, which compares favorably with top scoring model checkers on a particular class of problems, as previously characterized.
In the future  we intend to exploit internal information gathered by the sequential model checker (e.g., SeaHorn) to further speedup our results. We would also like to examine how to apply our approach to other hierarchies (e.g., trees).

\paragraph{Acknowledgement.}
This publication is part of a project that has received funding from the European Research Council (ERC) under the European Union's Horizon 2020 research and innovation programme (grant agreement No [759102-SVIS]).
    The research was partially supported by Len Blavatnik and the Blavatnik Family
    foundation, the Blavatnik Interdisciplinary Cyber Research Center, Tel Aviv
    University, and the United States-Israel
    Binational Science Foundation (BSF) grants No. 2016260 and 2012259.

\bibliographystyle{abbrv}
\bibliography{compositional-shorter}

\end{document}